\documentclass[12pt]{iopart}
\usepackage[utf8]{inputenc}
\usepackage[T1]{fontenc}

\usepackage{hyperref}
\hypersetup{
    colorlinks=true,
    linkcolor=blue,
    filecolor=magenta,      
    urlcolor=cyan,
    citecolor=blue}

\usepackage{graphicx}
\usepackage[justification=centering]{caption}
\usepackage[justification=centering]{subcaption}
\usepackage{siunitx}
\DeclareSIUnit\gauss{G}
\usepackage{booktabs}
\newcolumntype{C}{>{$}c<{$}}
\newcolumntype{L}{>{$}l<{$}}
\newcolumntype{R}{>{$}r<{$}}
\usepackage[noabbrev]{cleveref}
\usepackage{iopams}  
\usepackage{color}
\usepackage[dvipsnames]{xcolor}
\usepackage{todonotes}
\usepackage{cite}

\begin{document}

\title[EM PIC modeling of an ECR discharge in hydrogen]{Electromagnetic particle-in-cell modeling of an electron cyclotron resonance plasma discharge in hydrogen}

\author{D. Eremin$^1$, Yu. Sharova$^1$, L. Heijmans$^2$, A.M. Yakunin$^2$, M. van de Kerkhof$^{2,3}$, R.-P. Brinkmann$^1$, E. Kemaneci$^2$}

\address{$^1$ Institute of Theoretical Electrical Engineering, Ruhr University Bochum, Universitätsstrasse 150, D-44801 Bochum, Germany}

\address{$^2$ ASML Netherlands B.V., Veldhoven, The Netherlands}

\address{$^3$ Eindhoven University of Technology, Department of Applied Physics, Eindhoven, The Netherlands}

\ead{Denis.Eremin@rub.de}
\vspace{10pt}
\begin{indented}
\item[]
\today
\end{indented}

\begin{abstract}
A low pressure discharge sustained in molecular hydrogen with help of the electron cyclotron resonance heating at a frequency of 2.45 GHz is simulated using a fully electromagnetic implicit charge-
and energy-conserving particle-in-cell/Monte Carlo code. The simulations show a number of kinetic effects, and the results are in good agreement with various experimentally measured data such as electron density, electron temperature and degree of dissociation. 
%validating the code, in contrast to a fluid-based model incapable of reproducing such effects. 
%In particular, the electron energy distribution shows a non-Maxwellian shape resulting from a number of different electron heating mechanisms. 
The electron energy distribution shows a tri-Maxwellian form due to a number of different electron heating mechanisms, agreeing with the experimental data in the measured electron energy interval. The simulation results are also verified against a drift-diffusion model and proximity is observed between the computational results for the plasma density at the location of experimental measurement. However, the fluid approximation fails to accurately predict radical density and electron temperature because of the assumption of a single electron temperature. 
Special attention is paid to the characteristics of hydrogen radicals, whose production is strongly underestimated by the fluid model, whereas it is well predicted by the model considered here. The energy distribution of such radicals demonstrates the presence of a relatively large number of energetic hydrogen atoms produced by the dissociation of molecular hydrogen. The new insights are of significance for practical applications of hydrogen plasmas.  
\end{abstract}

%
% Uncomment for keywords
\vspace{2pc}


 %XXX
\section{Introduction} \label{sec1}

Extreme-Ultra-Violet (EUV) light of $92$ eV is essential for modern lithography tools, enabling the nanometer-sized-transistor manufacturing technology to meet the demand of high-quality integrated circuits within the semiconductor industry \cite{bakshi_2018}. The tools make use of a highly transparent hydrogen buffer gas to keep the multilayer mirrors clean of the carbonaceous layers \cite{feigl_2013} by igniting ``EUV-induced hydrogen plasma'' \cite{ven_2018, Beckers2019, Kerkhof2022}. Therefore, the construction materials of the lithography machines should endure the hostile plasma conditions. Preliminary tests on their durability are conducted in electron cyclotron resonance (ECR) discharges \cite{Bogaers2021} and their investigation is of the essence to provide the optimal testing conditions.

Depending on the description complexity, different numerical models are used, ranging from global models (e.g., \cite{hjartarson_2010,kemaneci_2017,kemaneci_2019, Kemaneci_2024}) taking into account only the volume-averaged characteristics and thus neglecting spatial resolution, fluid models (e.g., \cite{hagelaar_2004,rahimi_2014,Kemaneci_2015,Bogaers2021}) neglecting velocity space resolution, and kinetic models (usually of the particle-in-cell, or PIC type) accounting for the full phase space dynamics, kinetic and nonlocal effects. The latter are inherent to the ECR discharges due to the intrinsically non-Maxwellian electron distribution function owing to the anisotropic ECR electron heating \cite{Geller1996,Williamson1992,lieberman_2005} and collision mean free path being of the order of the system size for typical operation neutral gas pressures. PIC simulations have another advantage of being truly self-consistent, typically not requiring any external parameters or analytical models for the calculation of power absorbed in plasma, as is the case in, for example, fluid models. Unfortunately, conventional electromagnetic (EM) PIC models based on the explicit time integration schemes suffer from a serious numerical restriction limiting the time step to the time it takes a lightwave in vacuum to propagate through a computational grid cell \cite{birdsall_2005}
(it is worth noting that the conventional explicit electromagnetic PIC approach does not have another limit inherent for similar electrostatic codes which enforces
the cell size to be small compared to the Debye length \cite{Taccogna2023}). This is why reports of full-fledged EM PIC simulations of ECR plasmas used to be extremely scarce in the literature and are possible only at the expense of using a relatively small computational domain \cite{Gopinath95}. Usually, only PIC algorithms describing significantly reduced physics were published (e.g., \cite{Lampe1998,Takao2016,Fu2019,Nakamura2019}). A recent development in algorithms allowed to somewhat relax the time step restriction \cite{Porto2021,Porto2022}, but has not solved this problem completely. In this paper, we report the first attempt to apply an alternative method, the implicit energy- and charge-conserving PIC algorithm ECCOPIC2M-M \cite{Eremin2023} fully obviating the mentioned time step restriction, to modeling of an ECR discharge in hydrogen plasmas. 

The paper is organized as follows. The employed numerical model is described in Sec.\ref{sec2}. Sec.\ref{sec3} showcases comparison between the numerical results and experimental data measured previously, thereby validating the numerical model and the code resulting from its implementation. This section features some additional simulation data not accessible in the experiment that provide     
new insights and help to understand the physics of the discharge considered. The main conclusions are recapped in Sec.\ref{sec4}.  

%The corresponding high energy photons 

%Extreme Ultraviolet (EUV) plasma stands as a key component in cutting-edge technologies, particularly in the realm of semiconductor manufacturing and lithography. Its unique properties and behaviors present both opportunities and challenges for researchers and engineers alike. 

%Greater fidelity of PIC compared to COMSOL simulations. 

%%%%%%%%%%%%%%%%%%%%%%
%\clearpage
%%%%%%%%%%%%%%%%%%%%%%
\section{The model} \label{sec2}

The PIC/MCC model used in the present work is an expanded version of the energy- and charge-conserving PIC model used in \cite{Eremin2023}, where
additional modules for calculating the external magnetic field (see Appendix), treating the hydrogen plasma chemistry (see Sec.\ref{sec2.2}), and enabling the adaptive particle management (see Sec.\ref{sec2.3}) were implemented. The entire algorithm was parallelized on graphics processing units (GPUs) using a 2D generalization of the fine-sorting algorithm presented in \cite{mertmann_2011}.

\begin{figure}[h]
\centering
\includegraphics[width=10cm]{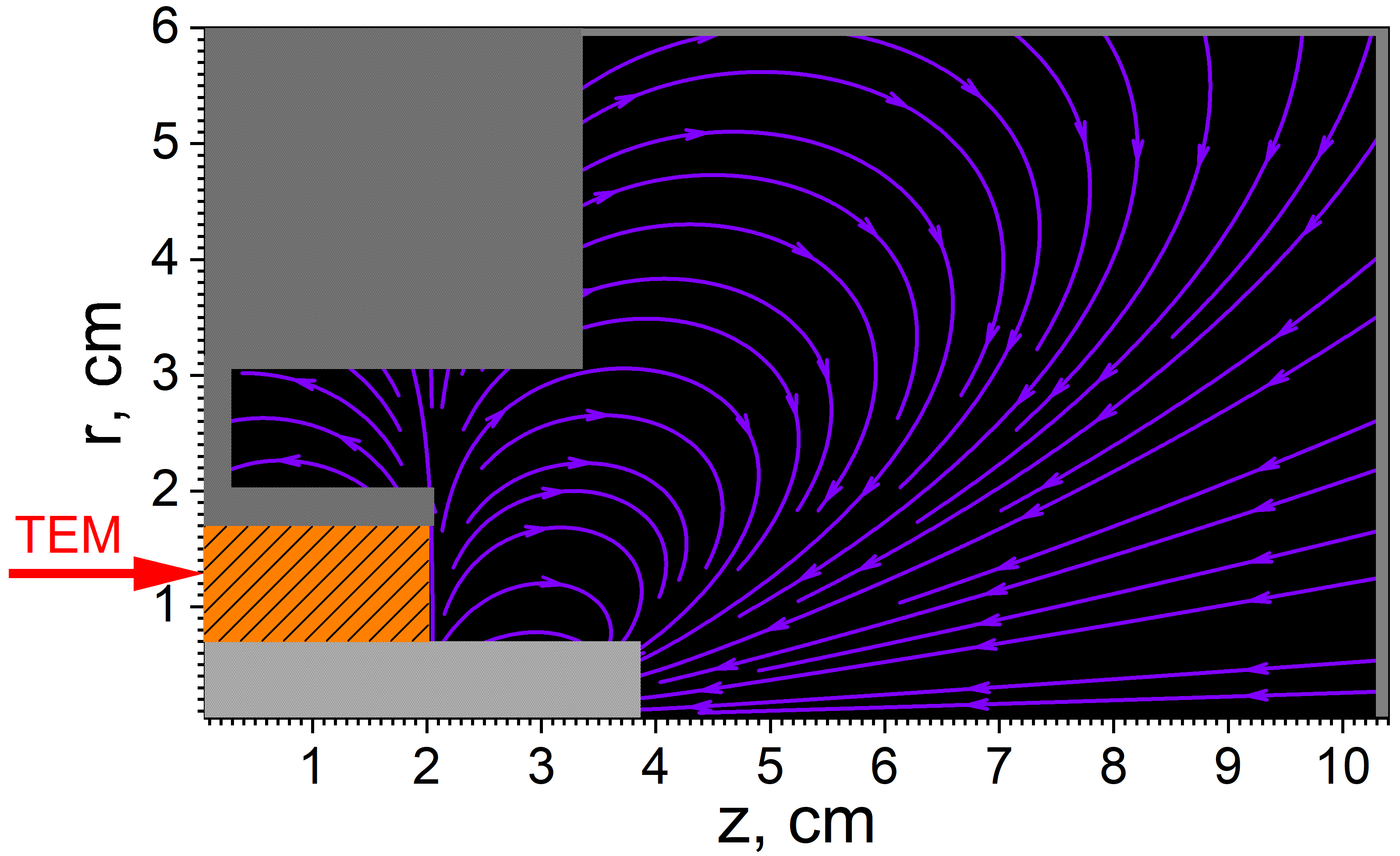}
\caption{Model geometry of the ECR discharge. The discharge is powered by the TEM wave entering through the dielectric alumina (orange). The central metal rod (light grey) contains permanent magnets generating the magnetic field. The dark grey area represents a metallic grounded surface. The cylindrical axis is pointing in the horizontal direction.}
\label{Fig1}
\end{figure}

\subsection{Electromagnetic particle-in-cell code} \label{sec2.1}

The geometry considered is based on the reduced setup studied in \cite{Bogaers2021}, which used the Aura-Wave ECR source \cite{Latrasse2017} to power the discharge. 
The model ECR discharge is sustained by a TEM electromagnetic wave entering through the dielectric port (see Fig.~\ref{Fig1}), the corresponding boundary condition to be described shortly. After propagating to the discharge chamber, the wave's electric field interacts with plasma electrons, which close to the central rod undergo a complex motion under the influence of the external magnetic field: bouncing along the magnetic field lines between the sheaths and/or areas of the enhanced magnetic flux density (mirror effect), rotating perpendicular to the magnetic field (electron cyclotron rotation), and executing slow drifts in the azimuthal direction due to the magnetic field line curvature and gradient of the magnetic flux density. The latter drifts are neglected in the model, which assumes azimuthal symmetry. Note that such a symmetry also eliminates a large number of instabilities
and the extraordinary mode (X-mode), which are expected to play a minor role in the considered discharge. Since the wave frequency of $2.45$ GHz can match the electron cyclotron frequency in certain regions, the electric field of the right-hand polarized wave (R-wave) can resonantly heat electrons there, increasing their velocity orthogonal to the magnetic field \cite{Popov1994, Geller1996, lieberman_2005}. If the velocity component parallel to the magnetic field is significant, it can modify the resonant condition through the corresponding Doppler shift.   

To self-consistently calculate propagation of the electromagnetic wave and sheath electric fields interacting with plasma particles on the time scale of the driving period and plasma diffusion to the reactor walls on much longer time scales, we used the implicit energy-conserving fully electromagnetic code ECCOPIC2M (albeit with an additional external magnetic field resulting in the suffix -M), which was described in detail in an earlier validation study \cite{Eremin2023}. The code is based on the implicit energy- and charge-conserving algorithm going back to the works \cite{markidis_2011,chen_2011} and a more recent adaptation to bounded and collisional technological plasmas in \cite{Eremin2022}. The code was verified with benchmarks \cite{turner_2013,charoy_2019,villafana_2021} and validated in \cite{Eremin2023,Berger2023}. The code also uses a nonuniform grid \cite{chacon_2013} to handle the radial region close to the axis by using the approach described in \cite{Eremin2023} and the axial region close to the top wall, where the plasma sheath is small.

Neglecting the azimuthal dependence and retaining only the field components related to a TM mode, which is assumed to dominate the physics, the field equations read

\begin{equation}
\left\{
\begin{array}{lll}
\mu \frac{\partial}{\partial t} H^\theta &=& \frac{1}{J}\left(\frac{\partial}{\partial\xi} E_\eta - \frac{\partial}{\partial \eta} E_\xi\right) \\
\left(\sigma+\epsilon \frac{\partial}{\partial t}\right) E^\xi &=& -j^\xi - \frac{1}{J}\frac{\partial}{\partial\eta} H_\theta \\
\left(\sigma+\epsilon \frac{\partial}{\partial t}\right) E^\eta &=& -j^\eta + \frac{1}{J}\frac{\partial}{\partial \xi} H_\theta,
\end{array}
\right. \label{eq2_1}
\end{equation}
and the field components are discretized according to the Yee scheme \cite{yee_1966}. In this equation, quantities with the superscript (subscript) denote the contravariant (covariant) field components \cite{chacon_2013}, $J_r = dr/d\xi$ with $\xi$ the logical radial coordinate, $J_z = dz/d\eta$ with $\eta$ the logical axial coordinate, the field grid in both logical coordinates being equidistant, and $J = J_rJ_z$. Note that $E_{\xi} = J_r^2 E^{\xi}$, $E_{\eta} = J_z^2 E^{\eta}$, and $H_\theta = r^2 H^\theta$.  

To complete the field equations, one needs to specify a form for the current density, which in the considered case of a two-dimensional problem are provided by \cite{chacon_2013}
\begin{equation}
\begin{array}{lll}
j^{\xi}_{s;i+1/2,j} &=& \frac{q_s}{J\Delta\xi\Delta\eta}\sum\limits_p w_p  v_{rp} J_r(\xi_p) S_{m-1}(\xi_p-\xi_{i+1/2}) S_m(\eta_p-\eta_j) \\ 
j^{\eta}_{s;i,j+1/2} &=& \frac{q_s}{J\Delta\xi\Delta\eta}\sum\limits_p w_p v_{zp} J_z(\eta_p) S_{m}(\xi_p-\xi_i) S_{m-1}(\eta_p-\eta_{j+1/2}),
\end{array} \label{eq2_2}
\end{equation}
where $\Delta \xi$ and $\Delta \eta$ denote the cell sizes of the logical grid, $S_m$ is the b-spline of order $m$ \cite{chen_2011}, $p$ subscript denotes contribution from the $p$th particle, $w_p$ is the superparticle weight (number of real particles in $p$th superparticle), $q_s$ is the charge of particle species $s$, and $i$ or $j$ denote the vertices of the corresponding field grid. Note that to ensure energy conservation, the same interpolation functions must be used for the electric field components \cite{markidis_2011,chen_2011}.

In order to make the model self-consistent, one needs to specify the time evolution of particle orbits. It is governed by the following equations in cylindrical coordinates (e.g., \cite{delzanno_2013b}),
\begin{equation}
\left\{
\begin{array}{rll}
m_s\frac{dv_r}{dt} &=& q_s\left(\frac{E_{\xi}}{J_r}+v_\theta B_{\eta,ext}\right) + m_s\frac{v_\theta^2}{r} \\
m_s\frac{dv_\theta}{dt} &=& q_s(v_z B_{\xi,ext} - v_r B_{\eta,ext}) -m_s\frac{v_r v_\theta}{r} \\
m_s\frac{dv_z}{dt} &=& q_s \left(\frac{E_\eta}{J_z} - B_{\xi,ext}\right)\\
\frac{d\xi}{dt} &=& \frac{v_r}{J_r}\\
\frac{d\eta}{dt} &=& \frac{v_z}{J_z} , 
\end{array}
\right.   \label{eq2_3}
\end{equation}
where particle velocities are $v_r = dr/dt$ and $v_z=dz/dt$ and the positions
are evolved in the logical coordinates \cite{chacon_2013,wang_1999,delzanno_2013}. Note that we assumed that the magnetic field of the wave is small compared to the external magnetic field created by the permanent magnets of the Aura-Wave source and neglected the former. 

Eqs. (\ref{eq2_1},\ref{eq2_2},\ref{eq2_3}) must be discretized in time. To this end, the time derivatives should be replaced by $\partial_t f = (f^{n+1}-f^n)/\Delta t$, and all other quantities should be taken at the mid-level between the new and the old time levels, $f^{n+1/2} = (f^n+f^{n+1})/2$. To integrate particle orbits in time, Eq. (\ref{eq2_3}) was used directly (in contrast to the typical use of the Boris's algorithm in cylindrical coordinates, e.g. \cite{mattei_2017}). This simplifies particle tracking and detection of particle crossings of cell boundaries, which is needed to ensure the charge conservation \cite{chen_2011}. 
%The use of such algorithm did not lead to noticeable errors or deviations from the results obtained with the usual Boris algorithm as was demonstrated, for example, in \cite{charoy_2019}.

The boundary condition at the dielectric is formulated in the form of the Mur condition for the outgoing TEM wave, $(J_z^{-1}\partial/\partial \eta - (\sqrt{\epsilon\mu}/c) \partial/\partial t)H_\theta^- = 0$ \cite{mur_1981}. Then, recalling that $H_\theta^-=H_\theta - H_\theta^+$ with $H_\theta$ the total magnetic field intensity and $H_\theta^+$ the incoming wave, one
can write the boundary condition for $H_\theta$ at the port as \cite{alvarez_2007,jimenez-diaz_2011,rahimi_2014}
\begin{equation}
\left(\frac{1}{J_z}\frac{\partial}{\partial \eta}  - \frac{\sqrt{\epsilon\mu}}{c}\frac{\partial}{\partial t} \right) H_\theta = S \label{eq2_5}
\end{equation}
with the source $S$ specified by the prescribed incoming wave, $S = -2 (\sqrt{\epsilon\mu}/c)\partial H^+_\theta/\partial t$. In this way the outgoing wave is allowed to leave the computational domain virtually without reflections at the port so that the radiation, which was not absorbed by plasma, is not accumulated in the discharge.

\subsection{Description of collisions: MCC module} \label{sec2.2}

The chosen chemistry set basically follows the one adopted in \cite{Astakhov2016}, albeit with some corrections. In particular, we have 
%added a new reaction R18 to account for the recombination and 
used a cumulative cross-section for the electron impact excitation dissociation (see below). Five species are tracked: electrons, three ion species
(H$^+$, H$^+_2$, and H$^+_3$), and two neutral species, atomic hydrogen H (aka hydrogen radical), and molecular hydrogen H$_2$. 
To model rotational excitations induced by an electron or ion impact, it is important to account for the rotational state populations of H$_2$, which we estimated using the corresponding formula given in \cite{simko_1997} and the approximate average temperature of H$_2$ obtained from experiment \cite{Bogaers2021}, which was about 350 K (0.3 eV). In addition, we assumed the background molecular hydrogen gas to have uniform density.
Main production of atomic hydrogen occurs through the dissociative excitation resulting from an electron impact (reaction R17). Up to about $20$ eV, this process is dominated by
excitation of the repulsive state $\mathrm{H_2}\,(b\, ^3\Sigma_{\mathrm{u}}^+)$ \cite{yoon_2008}, whereafter other processes start to contribute, such as dissociative excitation
with Lyman $\alpha$ radiation \cite{liu_1998}. We used cumulative cross-section from \cite{yoon_2008}. The threshold energies in reactions R28, R30-R33 were calculated from the energy diagram given in \cite{simko_1997} (see Fig.1 in the reference).
The negative threshold energies in R25 means that the corresponding reactions are exothermic and release energy, which we assumed to be transferred to the kinetic energy of the products, since the excited states were not tracked.   

Energetic electrons, which have gained energy from the electric field of the incoming wave through the ECR mechanism, produce new H$^+_2$ ions, mostly through the R2 reaction. The H$^+_2$ ions feed populations of other ion species through reactions R25 and R28, with the first reaction being the dominant and H$^+_3$ being the dominant ion species. H$^+_3$ can be converted back to H$^+_2$ via reactions R31 and R33, or converted to H$^+$ via reactions R30 or R32.

The Coulomb collisions (electron-electron and electron-ion) were included in the model, since it can be of the same order of magnitude as the electron elastic collision frequency with $H_2$ in the ECR region, where plasma density can be up to $\sim 10^{19}$ m$^{-3}$ (and both the cumulative Coulomb and the electron-neutral collision can be of the same order of magnitude for the neutral pressure of a few Pa). Furthermore, Coulomb collisions inherently drive the electron distribution to a Maxwellian, whereas electron-neutral collisions in general do not have that property. However, the contribution of Coulomb collisions rapidly diminishes away from the ECR region due to the steep fall of the electron density.

To model reactions listed in Tables \ref{tab:chemistry1} and \ref{tab:chemistry2} along with Coulomb collisions we employed the Monte Carlo binary collision algorithms \cite{Birdsall1991,Takizuka_1977}. 
In the implementation of all collisions except Coulomb ones we assumed that the background neutral gas density is uniform and much larger than all other densities, so that the corresponding density remains constant throught the simulation.

At the stainless steel walls (alumina), ions were assumed to recombine with the 100\%, and the hydrogen radical with the 10\% (0.1\%) \cite{hjartarson_2010} probability. Due to the relatively small plasma potential and the resulting sheath voltages \cite{Bogaers2021},
no secondary electrons were considered in the model.

\begin{table}[t!]
        \footnotesize
		\centering
		\begin{tabular*}{1.0\textwidth}{ l @{\extracolsep{\fill}} l l c c}
			\hline
			No.~ & reaction & process name & $\varepsilon_\mathrm{thr}, \mathrm{eV}$ &  src. \\
			\hline
			R1 & $\mathrm{e} + \mathrm{H_2} \rightarrow \mathrm{e} + \mathrm{H_2}$ & elastic scattering & - & \cite{brunger_2002} \\
			R2 & $\mathrm{e} + \mathrm{H_2} \rightarrow 2\mathrm{e} + \mathrm{H_2^+}$ & electron impact ionization & $15.43$ & \cite{janev2003} \\
			R3 & $\mathrm{e} + \mathrm{H_2} \rightarrow 2\mathrm{e} + \mathrm{H_2^+}\,(X\, ^2\Sigma_{\mathrm{g}^+}) \rightarrow 2\mathrm{e} + \mathrm{H^+} + \mathrm{H}$ & dissociative ionization & $18.46$ & \cite{janev2003} \\
			R4 & $\mathrm{e} + \mathrm{H_2} \rightarrow 2\mathrm{e} + \mathrm{H_2^+}\,(B\, ^2\Sigma_{\mathrm{u}^+}) \rightarrow 2\mathrm{e} + \mathrm{H^+} + \mathrm{H}$ & dissociative ionization & $32.69$ & \cite{janev2003} \\
			%\hline
			R5 & $\mathrm{e} + \mathrm{H_2}(v=0) \rightarrow \mathrm{e} + \mathrm{H_2}(v=1)$ & vibrational excitation & 0.58 & \cite{ehrhardt_1968} \\
		      R6 & $\mathrm{e} + \mathrm{H_2}(v=0) \rightarrow \mathrm{e} + \mathrm{H_2}(v=2)$ & vibrational excitation & 1.3 & \cite{ehrhardt_1968} \\
		      R7 & $\mathrm{e} + \mathrm{H_2}(v=0) \rightarrow \mathrm{e} + \mathrm{H_2}(v=3)$ & vibrational excitation & 1.5 & \cite{ehrhardt_1968} \\
			R8 & $\mathrm{e} + \mathrm{H_2}(j=0) \rightarrow \mathrm{e} + \mathrm{H_2}(j=2)$ & rotational excitation & 0.044 & \cite{england_1988} \\
		      R9 & $\mathrm{e} + \mathrm{H_2}(j=1) \rightarrow \mathrm{e} + \mathrm{H_2}(j=3)$ & rotational excitation & 0.073 & \cite{england_1988} \\
			R10 & $\mathrm{e} + \mathrm{H_2} \rightarrow \mathrm{e} + \mathrm{H_2}\,(B\, ^1\Sigma_{\mathrm{u}}^+)$ & electronic excitation & 11.3 & \cite{lxcat} \\
			R11 & $\mathrm{e} + \mathrm{H_2} \rightarrow \mathrm{e} + \mathrm{H_2}\,(C\, ^1\Pi_{\mathrm{u}})$ & electronic excitation & 12.4 & \cite{lxcat} \\
			R12 & $\mathrm{e} + \mathrm{H_2} \rightarrow \mathrm{e} + \mathrm{H_2}\,(a\, ^3\Sigma_{\mathrm{g}}^+)$ & electronic excitation & 11.8 & \cite{lxcat} \\
			R13 & $\mathrm{e} + \mathrm{H_2} \rightarrow \mathrm{e} + \mathrm{H_2}\,(c\, ^3\Pi_{\mathrm{u}})$ & electronic excitation & 11.72 & \cite{lxcat} \\
			R14 & $\mathrm{e} + \mathrm{H_2} \rightarrow \mathrm{e} + \mathrm{H_2}\,(d\, ^3\Pi_{\mathrm{u}})$ & electronic excitation & 14.0 & \cite{lxcat} \\
			R15 & $\mathrm{e} + \mathrm{H_2} \rightarrow \mathrm{e} + \mathrm{H_2}\,(e\, ^3\Sigma_{\mathrm{u}}^+)$ & electronic excitation & 13.0 & \cite{yoon_2008} \\
			R16 & $\mathrm{e} + \mathrm{H_2} \rightarrow \mathrm{e} + \mathrm{H_2}\,({\rm Rydberg, sum})$ & electronic excitation & 16.0 & \cite{lxcat} \\ 
			R17 \,\, & $\mathrm{e} + \mathrm{H_2} \rightarrow \mathrm{e} + 2\mathrm{H} $ & dissociation & 7.93 & \cite{yoon_2008} \\   
            %R18 \,\, & $\mathrm{e} + \mathrm{H_2^+} \rightarrow 2\mathrm{H} $ & dissociative recombination & -10.95 & \cite{janev2003} \\   
			\hline
			%30 & $\mathrm{He}^\ast + \mathrm{N}_2 \rightarrow \mathrm{He} + \mathrm{N\textsubscript{2}\textsuperscript{+}} + \mathrm{e}$ & Penning ionization & $5.0 \times 10^{-17}\,$m\textsuperscript{3}/s & (3)\\
			%31 & $\mathrm{He}^+ + \mathrm{He} + \mathrm{He} \rightarrow \mathrm{He\textsubscript{2}\textsuperscript{+}} + \mathrm{He}$ & ion conversion & $1.1 \times 10^{-43}\,$m\textsuperscript{6}/s & (3)\\
			%32 & $\mathrm{N_\textsubscript{2}\textsuperscript{+}} + \mathrm{N\textsubscript{2}} + \mathrm{He} \rightarrow \mathrm{N\textsubscript{2}\textsuperscript{+}} + \mathrm{He}$ & ion conversion & $1.9 \times 10^{-41}\,$m\textsuperscript{6}/s & (4)
		\end{tabular*}
		\caption{List of reactions between electrons and molecular hydrogen. The second column from the right indicates threshold energies $\varepsilon_\mathrm{thr}$.}
		\label{tab:chemistry1}
\end{table}

\begin{table}[t!]
        \footnotesize
		\centering
		\begin{tabular*}{1.0\textwidth}{ l @{\extracolsep{\fill}} l l c c}
			\hline
			No.~ & reaction & process name & $\varepsilon_\mathrm{thr}, \mathrm{eV}$ &  src. \\
			\hline
			R18 & $\mathrm{H^+} + \mathrm{H_2} \rightarrow \mathrm{H^+} + \mathrm{H_2}$ & elastic scattering & - & \cite{simko_1997} \\
			R19 & $\mathrm{H^+} + \mathrm{H_2} (j=0) \rightarrow \mathrm{H^+} + \mathrm{H_2}(j=2)$ & rotational excitation & 0.044 & \cite{simko_1997} \\
			R20 & $\mathrm{H^+} + \mathrm{H_2} (j=1) \rightarrow \mathrm{H^+} + \mathrm{H_2}(j=3)$ & rotational excitation & 0.077 & \cite{simko_1997} \\
			R21 & $\mathrm{H^+} + \mathrm{H_2}(v=0) \rightarrow \mathrm{H^+} + \mathrm{H_2}(v=1)$ & vibrational excitation & 0.516 & \cite{simko_1997} \\
			R22 & $\mathrm{H^+} + \mathrm{H_2}(v=0) \rightarrow \mathrm{H^+} + \mathrm{H_2}(v=2)$ & vibrational excitation & 1.0 & \cite{simko_1997} \\
			R23 & $\mathrm{H^+} + \mathrm{H_2}(v=0) \rightarrow \mathrm{H^+} + \mathrm{H_2}(v=3)$ & vibrational excitation & 1.5 & \cite{simko_1997} \\  
			R24 & $\mathrm{H^+} + \mathrm{H_2} \rightarrow \mathrm{H} + \mathrm{H_2^+}$ & charge transfer & 2.03 & \cite{simko_1997} \\   
			R25 & $\mathrm{H_2^+} + \mathrm{H_2} \rightarrow \mathrm{H_3^+} + \mathrm{H}$ & formation of $\mathrm{H_3^+}$ & -1.71 & \cite{phelps_1990} \\
			R26 & $\mathrm{H_2^+} + \mathrm{H_2} \rightarrow \mathrm{H_2} + \mathrm{H_2^+}$ & charge transfer & - & \cite{phelps_1990} \\
			R27 & $\mathrm{H_2^+} + \mathrm{H_2}(v=0) \rightarrow \mathrm{H_2^+} + \mathrm{H_2}(v=1)$ & vibrational excitation & 0.516 & \cite{phelps_1990} \\
			R28 & $\mathrm{H_2^+} + \mathrm{H_2} \rightarrow \mathrm{H^+} + \mathrm{H} + \mathrm{H_2}$ & dissociation & 2.65 & \cite{phelps_1990} \\   
			R29 & $\mathrm{H_3^+} + \mathrm{H_2} \rightarrow \mathrm{H_3^+} + \mathrm{H_2}$ & elastic scattering & - & \cite{simko_1997} \\ 
			R30 & $\mathrm{H_3^+} + \mathrm{H_2} \rightarrow \mathrm{H_2} + \mathrm{H} + \mathrm{H^+} + \mathrm{H}$ & dissociative charge transfer & 8.84 & \cite{peko_1997} \\ 
			R31 & $\mathrm{H_3^+} + \mathrm{H_2} \rightarrow \mathrm{H_2} + \mathrm{H} + \mathrm{H_2^+}$ & dissociative charge transfer & 6.19 & \cite{peko_1997} \\ 
			R32 & $\mathrm{H_3^+} + \mathrm{H_2} \rightarrow \mathrm{H^+} + \mathrm{H_2} + \mathrm{H_2}$ & dissociation & 4.36 & \cite{peko_1997} \\ 
			R33 & $\mathrm{H_3^+} + \mathrm{H_2} \rightarrow \mathrm{H_2^+} + \mathrm{H} + \mathrm{H_2}$ & dissociation & 6.19 & \cite{phelps_1990} \\ 
			R34 & $\mathrm{H} + \mathrm{H_2} \rightarrow \mathrm{H} + \mathrm{H_2}$ & elastic scattering & - & \cite{simko_1997} \\ 
			%R36 & $\mathrm{H_2} + \mathrm{H_2} \rightarrow \mathrm{H_2} + \mathrm{H_2}$ & elastic scattering & - & \cite{simko_1997} \\   
			\hline
		\end{tabular*}
		\caption{List of reactions between heavy particles and molecular hydrogen. The second column from the right indicates threshold energies $\varepsilon_\mathrm{thr}$.}
		\label{tab:chemistry2}
\end{table}

\subsection{Adaptive particle management for an energy-conserving PIC algorithm} \label{sec2.3}

Due to cylindrical geometry and an equidistant field grid the superparticle weight is initially proportional to its radial position if
an approximately uniform number of superparticles per cell is used. In the course of a simulation light superparticles at the center tend to be replaced
by few heavy particles from the radial periphery, which leads to excessive statistical noise at the center \cite{Wang2010}. Lightweight superparticles should therefore be merged when
they go toward larger radial locations and the heavy superparticles should be split when they go towards the discharge radial center. 
%Simultaneously, binary collision algorithm for superparticles with unequal weights (see the previous section) must be counteracted with a merging algorithm reducing the superparticle number. 
Furthermore, the ECR discharge considered is highly nonuniform, featuring a large plasma density close to the resonance location, with the density rapidly falling toward the bulk. To ensure a good phase space resolution it is advisable to split superparticles diffusing from the ionization zone toward discharge walls. The spliting
and merging algorithms form the basis for the adaptive particle management \cite{Lapenta1994,Assous2003,Welch2007,Teunissen2014,Vranic2015}. Such algorithms should obviously keep the numerical model similar to the one before the splitting or merging. It is therefore reasonable to ensure the conservation of kinetic energy, momentum, grid moments, and minimal perturbation of the local particle distribution in phase space \cite{Welch2007,Vranic2015}. The splitting algorithm is trivial to build for a collisional plasma, since one can simply split a superparticle into several parts, which will naturally be randomly scattered by the collision algorithm as the time goes by so that phase space is sampled better. The merging algorithm is much more complicated and needs to be constructed with caution.   

To handle this issue for an energy-conserving algorithm, we have implemented an algorithm merging $N$ superparticles with $N>2$ into two. The grid moment relevant for a charge-conserving electromagnetic code is the charge density. If the merging algorithm is invoked after the coupled energy-conserving integration of particle orbits and fields over the PIC cycle time $\Delta t$ and it does not produce any spurious charge density in a cell at grid locations relevant for the interpolation scheme chosen, the electric field discretized on the grid also remains intact. Provided that particle merging ensures energy conservation, the total energy, equal to a sum of the field energy, which can be expressed through the field values discretized on the grid, and of the kinetic particle energy, which can be expressed as the sum of the kinetic energies of individual superparticles, is exactly conserved as well. In this particular work, we have considered $m=1$ in the interpolation functions used in Eq.(\ref{eq2_2}). It follows that the pertinent charge density interpolation function is $S_1(r)S_1(z)$, also known as cloud-in-cell (CIC) \cite{birdsall_2005}. In the 2D geometry considered, a particle contributes to $4$ vertices of the cell it resides in, where for each of the vertices the deposited charge can be calculated as $Q_{ij} = \sum_{p=0}^{N-1}q w_p S_r(r_p-r_i)S_z(z_p-z_j)$ with $i=l,h$, where $l={\rm int}(r_p)$, ${\rm int}$ denotes the round-off to the next integer from the low side, $h=l+1$, and a similar treatment for $j$. If all $Q_{ij}$ are to be kept unchanged after the merge, one needs to guarantee that
$Q_{ij} = q W [S_r(r_0-r_i)S_z(z_0-z_j)+S_r(r_1-r_i)S_z(z_1-z_j)]$, where we used an assumption that both new superparticles have an equal weight $W=(\sum_{p=0}^{N-1}q w_p)/2$ (Note that this explicitly excludes a possibility of new weights being small as could result from the algorithm suggested in \cite{Welch2007}). Because of $\sum_{i,j} S_r(r_p-r_i)S_z(z_p-z_j)=1$ only three $Q_{ij}$ are independent. This results in three equations for four unknown coordinates, which for the linear interpolation function $S$ can be formulated as
\begin{equation}
\begin{array}{lll}
&&  q_0 = h_{r0} + h_{r1} \\
&&  q_1 = h_{z0} + h_{z1} \\
&&  q_2 = h_{r0}h_{z0} + h_{r1}h_{z1},
\end{array} \label{eq2_6}
\end{equation}
where $h_{r,z} = (\{r,z\}_p - {\rm int}(\{r,z\}_p))/\Delta \{r,z\}$, 
$q_0 = 2(Q_{hh}+Q_{hl})$, $q_1 = 2(Q_{hh}+Q_{lh})$, and $q_2 = 2Q_{hh}$. One can therefore formulate the problem of finding a free parameter, e.g., $0\leq h_{r0}<1$, that guarantees that $0\leq \{h_{r1},h_{r1},h_{r1}\}<1$. Taking into account Eqs.(\ref{eq2_6}) results in
\begin{equation}
\begin{array}{lll}
&&  h_{r1} = q_0 - h_{r0} \\
&&  h_{z0} = q_1 + \frac{q_2-q_1h_{r0}}{2h_{r0}-q_0} \\
&&  h_{z1} = \frac{q_1 h_{r0} - q_2}{2h_{r0}-q_0},
\end{array} \label{eq2_7}
\end{equation}
where $h_{r0}$ should be uniformly sampled from 
$${\rm max}(0,q_0-1) \leq h_{r0} \leq {\rm min}\left(\frac{q_0}{2}, \frac{q_2}{q_1}, \frac{q_0-q_1}{2-q_1}, -\frac{q_2}{q_1}+q_0, -q_0 + \frac{q_2+q_0}{q_1} \right)
$$ 
and 
$${\rm max}\left(\frac{q_0}{2}, \frac{q_2}{q_1}, \frac{q_0-q_1}{2-q_1}, -\frac{q_2}{q_1}+q_0, -q_0 + \frac{q_2+q_0}{q_1} \right)
\leq h_{r0} \leq q_1.
$$

Further, to make sure that the particle distribution function is not significantly perturbed, only particles with similar velocities should be chosen for merging. This is easiest achieved if one restricts the velocity interval for the particles to be merged to the velocities of the order of the thermal velocity, where particles are most abundant. Then, dividing the interval into $N$ bins for each of the velocity components, one can search for another three particles in the same grid cell with each of the velocity components falling into the same bin. Since the code is parallelized on GPU, such a search is performed by a separate thread for each of the particles, resulting in a tolerable computation cost. Then, the chosen four particles are merged into two. The race condition is avoided with help of the locking mechanism, and to mitigate the jamming we used $r<0.25$ throttle with $r$ the uniformly distributed pseudorandom number. This problem can be further alleviated for neutral atomic hydrogen species, where for a 2D problem one can focus on merging particles with close radial and azimuthal velocity components. To this end, one can create ordered lists of particle indices that belong to a particular bin. Then, one can eliminate the race condition altogether by choosing successive batches of particles belonging to the same two-dimensional bin. Such an algorithm is particularly efficient.

When merging particles into two, it is possible to retain conservation of the cumulative momentum and energy in each of the coordinates. It is straightforward to show that this can be achieved if in each coordinate $\alpha$ final velocities of the particles satisfy the following formulas:
\begin{equation}
v^\alpha_{1,2} = \frac{P^\alpha}{M} \pm \left(\frac{2E^\alpha}{M} - \frac{(P^\alpha)^2}{M^2}  \right)^{1/2} \label{eq2_8}
\end{equation}
with $M$ the mass of the resulting particles, and $P$ and $E$ cumulative momentum and energy of the original particles, respectively.

\subsection{Time-slicing algorithm for the treatment of atomic hydrogen} \label{sec2.4}

Establishment of the atomic hydrogen density profile takes time which is about two orders of magnitude larger compared
to analogous times for the plasma species. To manage this problem, one can resort to the ``time-slicing'' method \cite{Kushner2009}, where
the plasma and atomic hydrogen species are evolved on separate time scales, with the time step for the atomic hydrogen
being $100$ times larger than the time step for the plasma species. The underlying ansatz is that to calculate
sources of hydrogen atoms (with the dominant source coming from the dissociation reaction R17) one needs to know only the electron velocity distribution function, which in general can be thought of instantaneously adjusting to the slowly changing neutral background. Since in the first approximation in the considered problem we neglected the reciprocal influence of the atomic hydrogen on the discharge properties, one can also view it as the simpler test particle method and implement a model of the atomic hydrogen transport in a separate code adopting prefixed production rate \cite{Trieschmann2015}, e.g., calculated from the PIC/MCC code. However, in the given work the atomic hydrogen model was incorporated in the PIC/MCC code in such a way that for every dissociation reaction initiated by an electron impact there were created 100 pairs of hydrogen atoms with final velocities after the dissociation sampled randomly using the Monte Carlo procedure respecting the energy conservation. To get a smoother atomic hydrogen density profile, the initial positions of the new hydrogen atoms were shifted from the dissociation event location by a distance proportional to their velocities and $R\Delta t$ with $R$ random number uniformly distributed from $0$ to $1$. 
The hydrogen atoms were then moved and elastically scattered with their own time step. 

\section{Results} \label{sec3}

To make sure that the constructed code faithfully reproduces the real physics, we validated the code results 
%\ek{The simulation results are bench-marked} 
against the corresponding measurements done in \cite{Bogaers2021} at the $p=5$ Pa molecular hydrogen pressure and a power variation, which is the subject of Sec.\ref{sec3.1}. Having validated the code with the available experimental data, we discuss 
some insights made purely on the basis of the code output data in Sec.\ref{sec3.2}. In particular, we address the electron heating and energy distribution of the atomic hydrogen, along with its density profile.
%\ek{the electron heating and energy distribution of the atomic hydrogen, along with its density profile.}

\subsection{Validation against measurements} \label{sec3.1}

To validate the ECCOPIC2M-M code we chose to compare electron density, electron temperature, and the molecular hydrogen dissociation fraction between experimental measurements and the code output. Due to the relatively large computational cost of the ECCOPIC2M-M simulations, only four different power values were sampled (25 W, 67 W, 96 W, and 125 W).  

\begin{figure}[h]
\centering
\includegraphics[width=10cm]{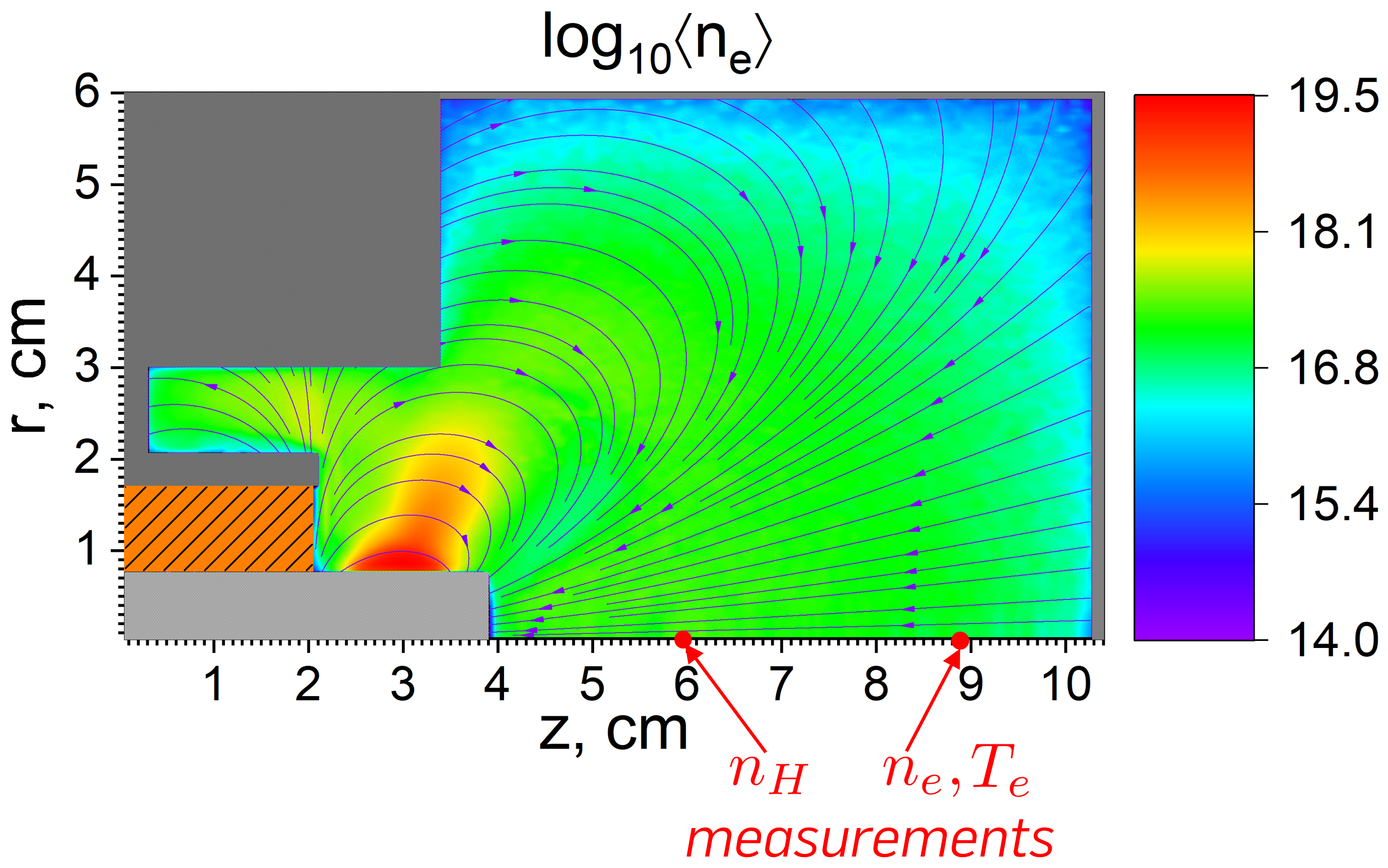}
\caption{Logarithm of the period-averaged electron density (measured in m$^{-3}$) for the P=125 W run with locations corresponding to the
electron temperature and density, and the atomic hydrogen density measurements in the experiment.}
\label{ne_and_measurement_locations}
\end{figure}

Fig.\ref{ne_and_measurement_locations} shows the logarithm of the period-averaged electron density with field lines of the magnetic field produced by the permanent magnets. One can observe a dense electron
population sustained close to the central rod ($r<1$ cm). This population has an arc-like boundary closely following the local magnetic field. Another electron population has a tongue-shaped form and spans a radial interval ($1<r<2.5$ cm). The corresponding electron density shape peaks close to the location of the magnetic field minimum on each magnetic field line, which can be explained by the fact that the ECR heating produces a large number of electrons with large pitch angles, which are efficiently confined by the magnetic field mirrors. As will be seen in Sec.\ref{sec3.2}, these two electron populations appear to be generated by two different ionization sources related to different energetic electron populations.
Similar electron density profiles were obtained for all powers, so that only one case is shown. Fig.\ref{ne_and_measurement_locations} also indicates experimental measurement points for electron density and temperature along with atomic hydrogen density. The corresponding quantities were also extracted from the simulation output at the shown locations. Note that the measurements were made in the ``quiescent'' region on the axial axis, away from the ``heating'' region, where electrons absorb energy from the electromagnetic fields and produce new electrons via ionization processes, and where most electrons are confined. 

\begin{figure}[h]
\centering
\includegraphics[width=8cm]{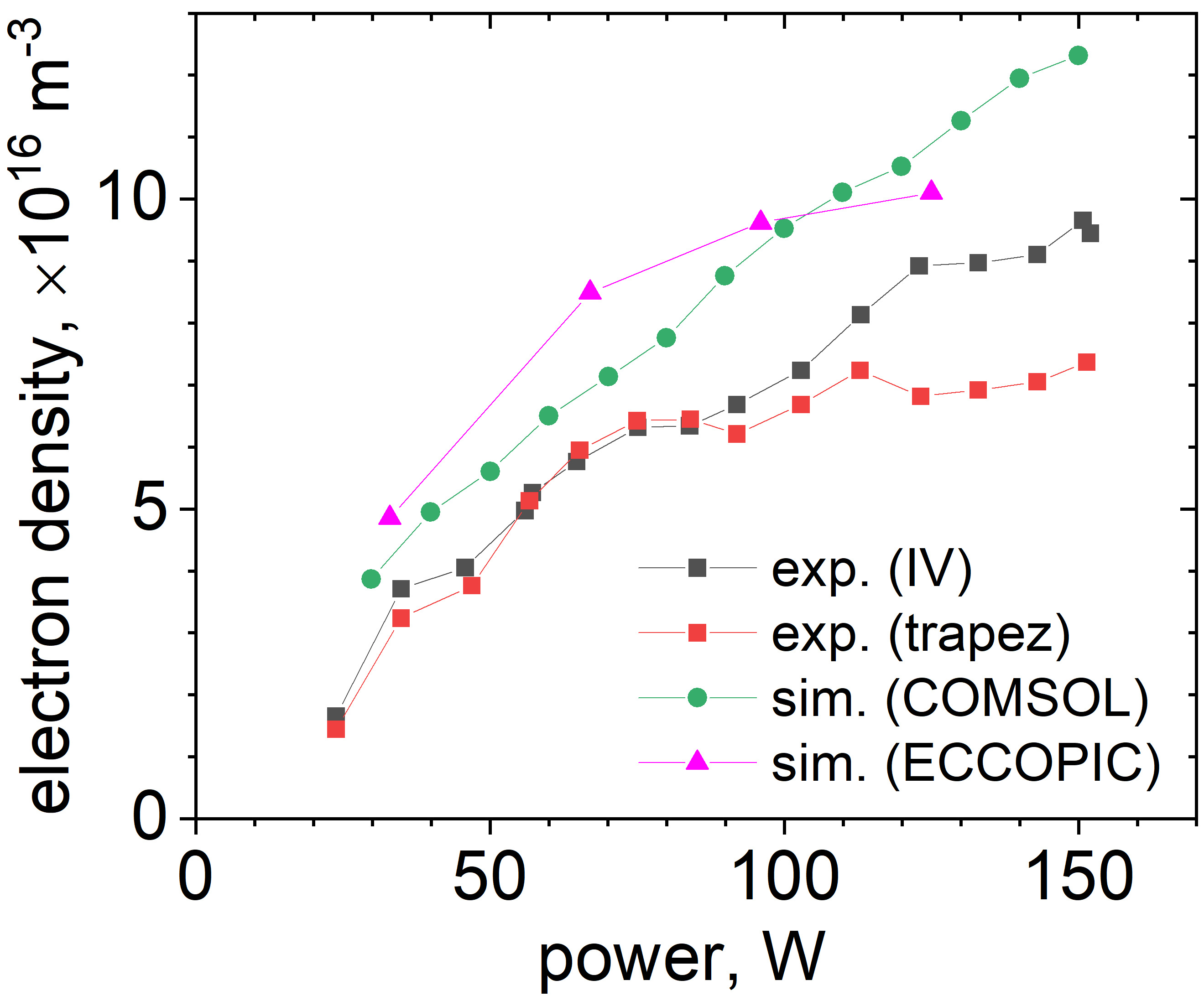}
\caption{Electron density variation with power obtained from two different experimental diagnostics (black and red) and from modeling, using the fluid-based COMSOL (green) and PIC/MCC-based ECCOPIC2M-M (magenta) codes.}
\label{Fig_denEl_vs_power}
\end{figure}

Fig.\ref{Fig_denEl_vs_power} shows a good agreement of the electron density provided by the ECCOPIC2M-M and Langmuir probe-based experimental diagnostics. Although COMSOL electron density values demonstrate good agreement as well, they tend to exhibit stronger growth with increasing power compared to the experiment.

\begin{figure}[h]
\centering
\includegraphics[width=8cm]{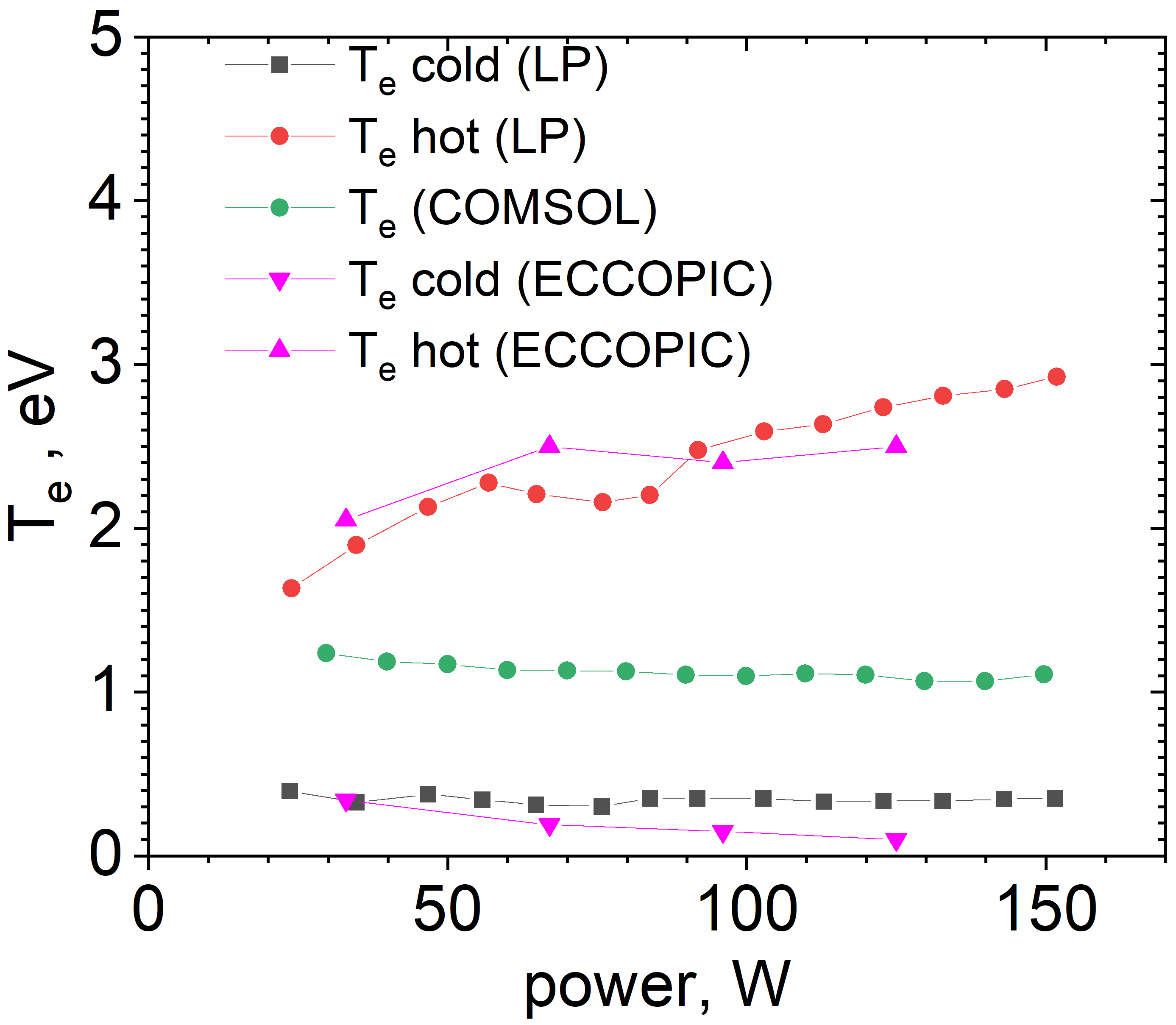}
\caption{Electron temperatures extracted from the bi-Maxwellian EEPF obtained from the experiment (black) and from modeling, using the fluid-based COMSOL (green) and PIC/MCC-based ECCOPIC2M-M (magenta) codes.}
\label{Fig_Te_vs_power}
\end{figure}

In the experiment, it was possible to reliably measure the electron energy distribution function (EEDF) up to energies of $12$ eV using the Langmuir probe, which exposed the bi-Maxwellian character of the EEDF in this pressure range (see Fig.6.15 in \cite{Bogaers2021}), with a cold low-temperature and a high-temperature energetic tail electron population. 
A plot of the corresponding temperatures estimated from the experiment \cite{Bogaers2021} and the ECCOPIC2M-M code is shown in Fig.\ref{Fig_Te_vs_power}. Evidently, the hot temperature is reliably reproduced by the code, whereas the low temperature seems to exhibit a different trend from the experiment, somewhat falling down with power. We assume that this can be related to the reduced 2D model geometry used in the code, which excludes many possible instablities potentially excited in ECR plasmas with growing power \cite{Geller1996} that can heat electrons. In contrast, the COMSOL fluid-based model yields only a single temperature and so is not capable of faithfully reproducing the discharge physics in this respect. 

\begin{figure}[h]
\centering
\includegraphics[width=8.7cm]{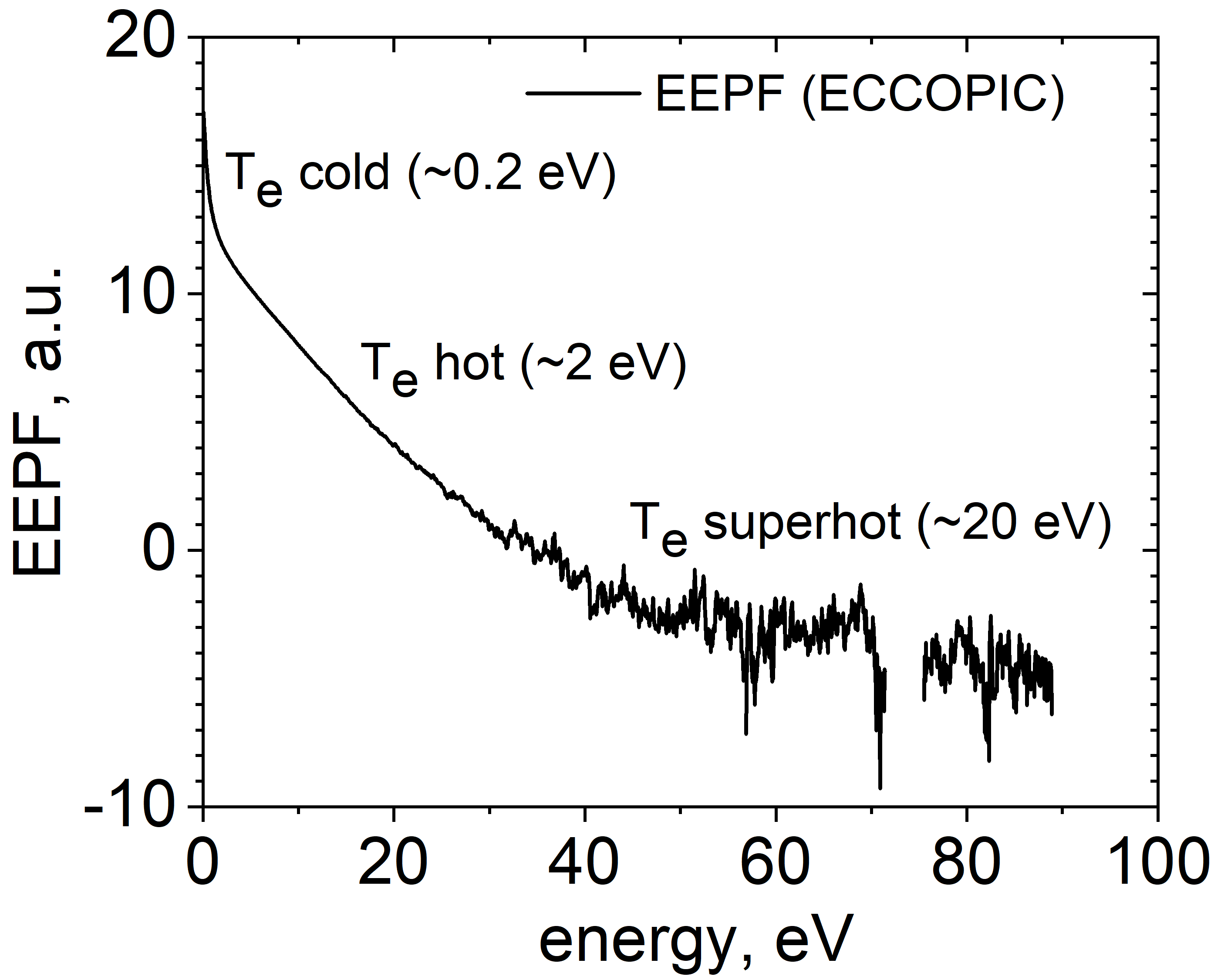}
\caption{Electron energy probability function provided by PIC/MCC simulation with P=67 W exhibits three electron populations (cold, hot and superhot) with distinct characteristic temperatures.}
\label{Fig_EEPF_67W}
\end{figure}

On closer scrutiny, the electron energy probablity function (EEPF) obtained from the PIC/MCC simulations reveals that, when viewed on a larger energy scale, there seems to be an additional, ``superhot'' energetic electron population present at high energies approximately above $40$ eV (see Fig.\ref{Fig_EEPF_67W}). The emergence of the third group of electrons can be linked to the different electron heating mechanisms present in the discharge (see Sec.\ref{sec3.2}). However, a detailed study of these mechanisms goes beyond the scope of the present work.

\begin{figure}[h]
\centering
\includegraphics[width=9cm]{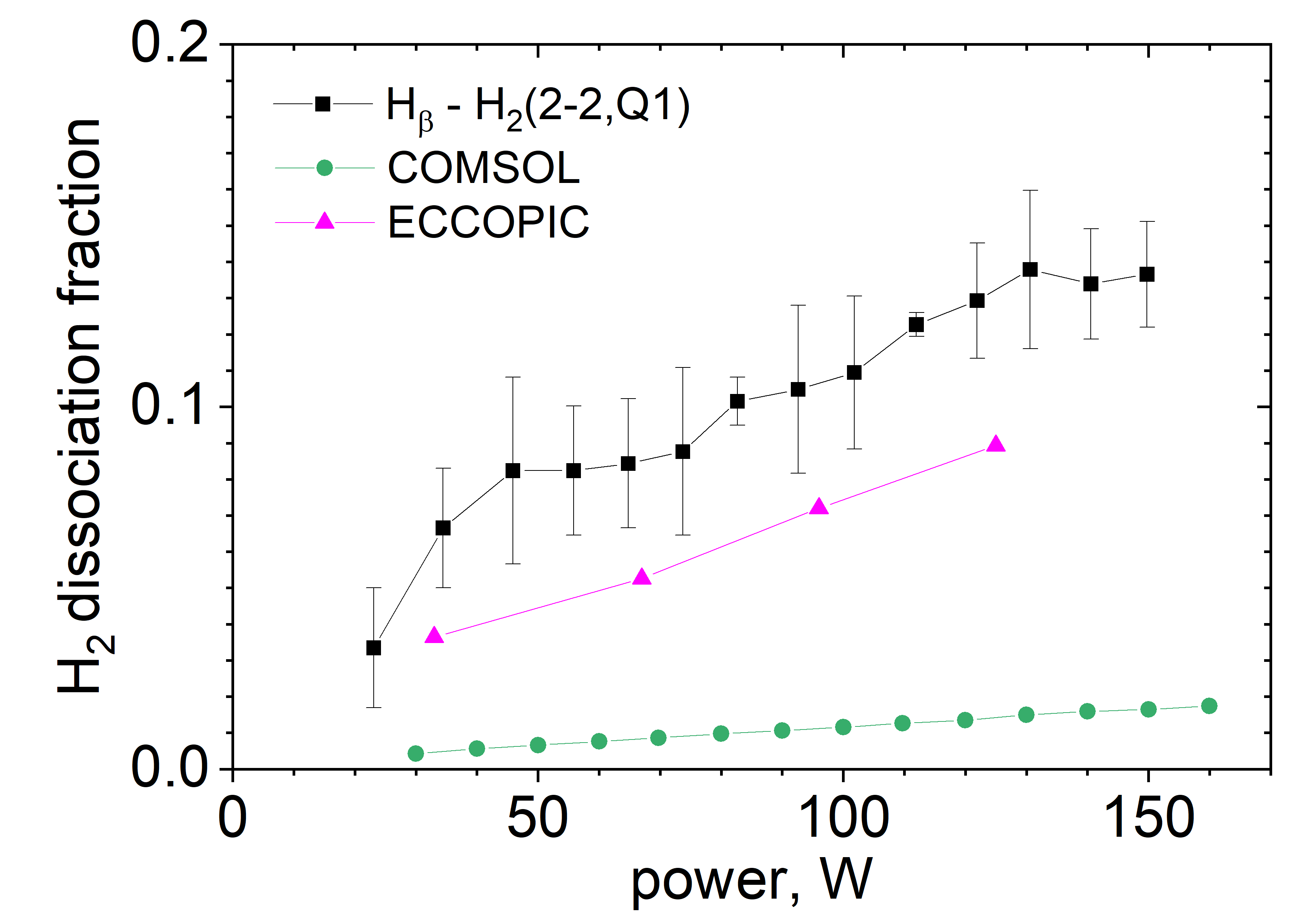}
\caption{Dissociation fraction of the molecular hydrogen comparison between an experimental diagnostics (black), COMSOL (green), and ECCOPIC2M-M (magenta) simulations.}
\label{Fig_H2_dissociation_fraction_vs_power}
\end{figure}

As aforementioned, the atomic hydrogen species is believed to be potentially significant for the reactor environment. Therefore, to validate the ECCOPIC2M-M code we included the molecular hydrogen dissociation fraction which is intimately related to the atomic hydrogen density and is given by
\begin{equation}
X_{diss} = \frac{0.5 n_H}{0.5 n_H + n_{H_2}} \label{eq3_1}
\end{equation}
The corresponding data comparison between the experiment and different numerical models is shown in Fig.\ref{Fig_H2_dissociation_fraction_vs_power}. Of all the experimental data given in \cite{Bogaers2021}, only the data provided by the $H_\beta$ line from the $H_2$ actinometry diagnostics were chosen, since they are the most reliable \cite{Bogaers2021}. The plot demonstrates that the PIC/MCC model leads to much better agreement with the experimental data compared to the fluid model. This can be explained by the fact that the production of atomic hydrogen is dominated by the dissociation reaction R17 triggered by the energetic electrons. 
%Since COMSOL does a poor job in describing such electrons, it dramatically underestimates production of atomic hydrogen species. 
Since the fluid approximation ignores the energetic electrons, it dramatically underestimates the production of atomic hydrogen species.

\subsection{Additional diagnostics} \label{sec3.2}

Particle-in-cell codes provide a plethora of various information going far beyond the data discussed in the previous section. In this section, we analyze some of it in the example of data pertaining to the simulation with P=125 W.

\begin{figure}[h]
\centering
\includegraphics[width=16cm]{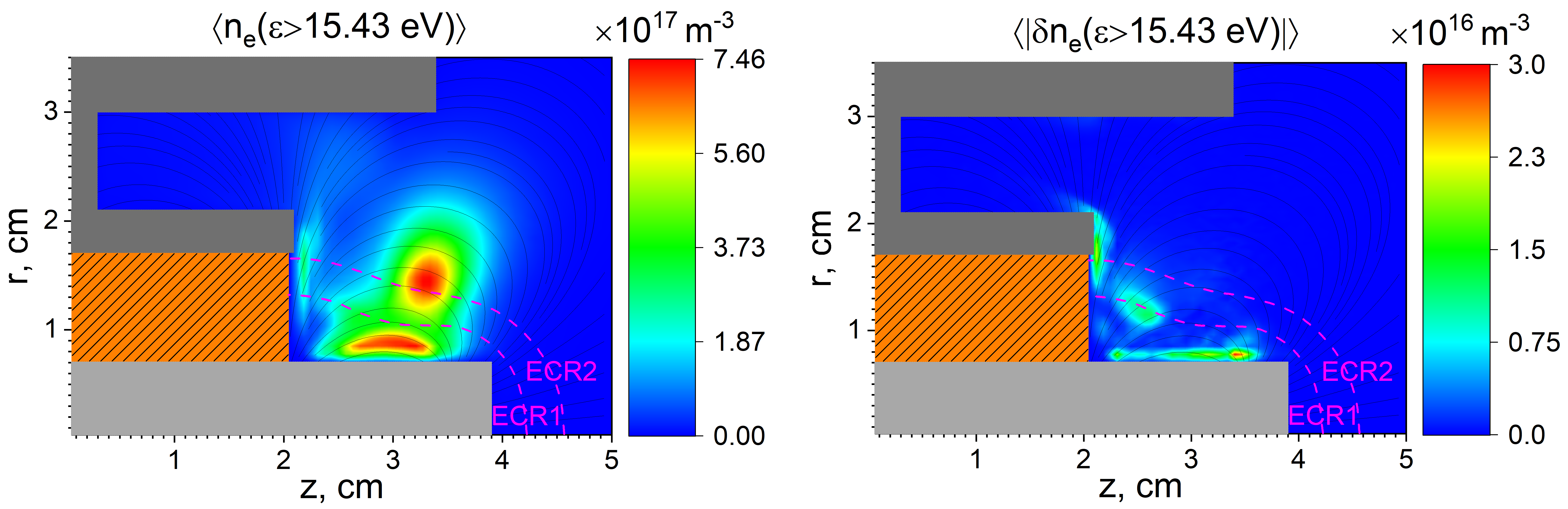}
\caption{Period-averaged density of energetic electrons (left) and its oscillation level (right).}
\label{denEnEl}
\end{figure}

The electron density profile shown in Fig.\ref{ne_and_measurement_locations} is the result of electron transport to the walls where they recombine
and the ionization processes (predominantly the reaction R2 with the lowest energy threshold) that replenish the transport losses. It is therefore
instructive to analyze the population of energetic electrons with the energy above the threshold of reaction R2. The plot exhibits two peaks, one close to the central rod, and the other away from it (see Fig.\ref{denEnEl}, left; Note that to better resolve the important region where the main activity takes place, only this region is plotted on this and the next two plots). It might seem that the second energetic electron population is located on the resonant surface corresponding to the second ECR harmonic (where $B_{ECR2} = m_e\omega/2e$), where ECR heating is also possible \cite{Popov1994}. However, it turns out that this electron population consists mainly of electrons heated by the fundamental ECR harmonic (where $B_{ECR1} = m_e\omega/e$). To see that, we use the fact that at the stationary phase, when the period-average value of any quantity should be zero, one can estimate the oscillation level of that quantity from   
\begin{equation}
\left< |\delta Q| \right > = \frac{1}{T} \int\limits_0^T |Q(t)-\overline{Q}|, dt \label{eq3_2}
\end{equation}
which for the energetic electron density results in Fig.\ref{denEnEl}, right. This plot shows that energetic electrons are produced in different regions
compared to the peaks of their period-averaged density. One can identify three main regions: two regions close to the conducting surfaces and one region detached from any surface. Although in this paper we do not intend to elaborate on particular details of the electron heating mechanisms, one can assume that the electron heating close to the surfaces is caused by the sheath motion-related electron heating \cite{Godyak1972,LiebermanGodyak1998} modified by the presence of magnetic field \cite{Eremin2023b}. The electron heating/energization in the third group occurs due to the ECR heating at the fundamental harmonic. Energetic electrons from this group then move along the magnetic field lines and, due to the large pitch angles resulting from the ECR heating, are efficiently confined by the magnetic mirrors, feeding the energetic electron population having density peaked at $1.4 < r < 1.7$ cm in Fig.\ref{denEnEl}, left. 
\begin{figure}[h]
\centering
\includegraphics[width=16cm]{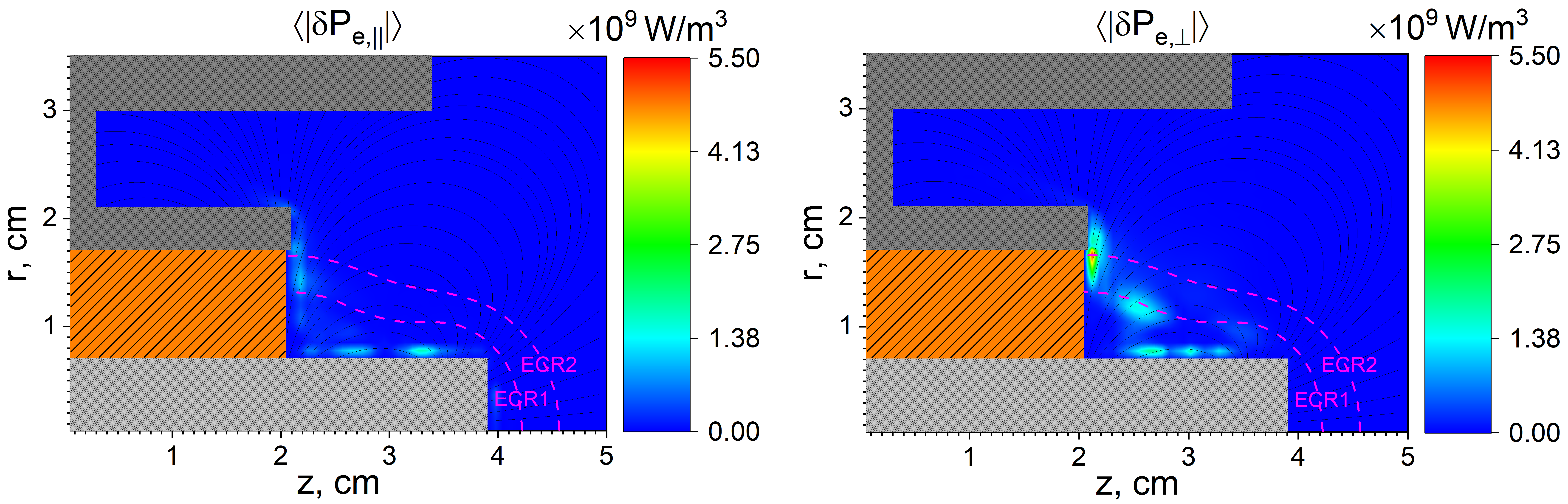}
\caption{Oscillation levels of the power density absorbed by electrons parallel (left) and perpendicular (right) to the background magnetic field.}
\label{deltaPowEl}
\end{figure}
This is further corroborated in Fig.\ref{deltaPowEl}, where oscillation levels of electron power absorption parallel and perpendicular to the background magnetic field are shown. Here, $P_{e,\parallel} = j_{e,\parallel}E_{e,\parallel}$ and $P_{e,\perp} = {\bf j}_{e,\perp}\cdot{\bf E}_{e,\perp}$, where ${\bf v}_\parallel = v_\parallel {\bf b}$ with ${\bf b}= {\bf B}_{ext}/B_{ext}$ and $v_\parallel = {\bf b}\cdot{\bf v}$, and ${\bf v}_\perp = {\bf v}-{\bf v}_\parallel$ for an arbitrary vector ${\bf v}$. One can clearly see that the ECR heating indeed takes place only perpendicular to the magnetic field as it should be, whereas electron heating close to the electrodes shows a more complicated pattern. Close to the central rode, where the magnetic field is relatively strong, the electron heating is reminiscent of the patterns observed in \cite{Eremin2023b} for an rf-magnetron, which involved the mirror-effect heating (MEH), magnetized bounce heating (MBH), and the Hall heating mechanisms (for the latter, see also \cite{zheng_2019,Eremin2023a}). The electron heating pattern close to the grounded conducting surface at $1.7 < r< 2.1$ cm is likely to be dominated by the Hall heating due to the almost parallel magnetic field to the electrode surface there.

\begin{figure}[h]
\centering
\includegraphics[width=8.5cm]{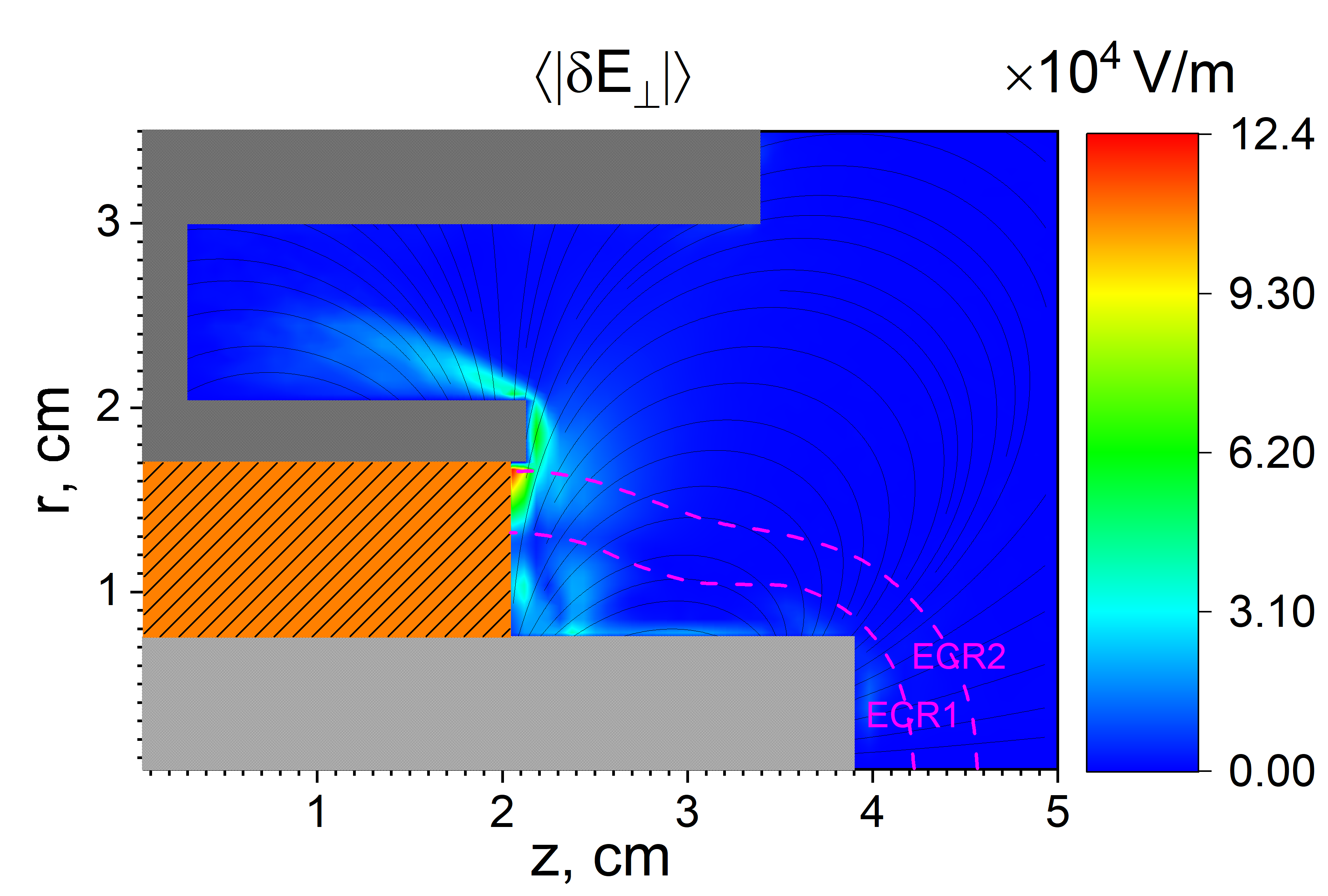}
\caption{Oscillation level of the electric field perpendicular to the background magneic field.}
\label{deltaEPerp_av_125W}
\end{figure}

An interesting observation is that the perpendicular electric field seems to be suppressed in the ECR resonance area, see Fig.\ref{deltaEPerp_av_125W}. This is consistent with observation made in other works (see, e.g., \cite{Khokar2018}) and is due to the strong absorption of the electromagnetic waves transferring their energy to electrons.

\begin{figure}[h]
\centering
\includegraphics[width=10cm]{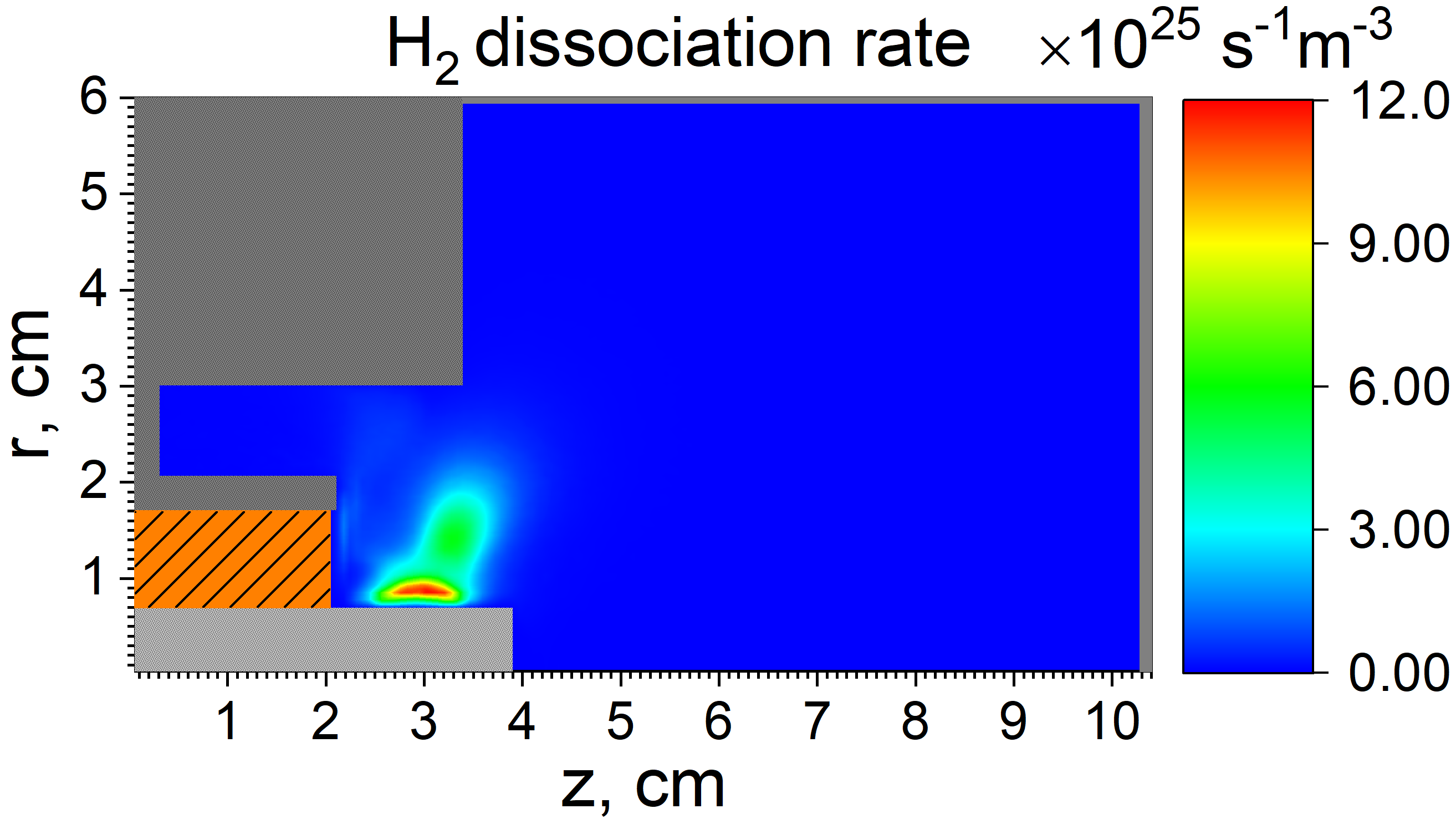}
\caption{Dissociation rate of the molecular hydrogen.}
\label{Fig2H}
\end{figure}

\begin{figure}[h]
\centering
\includegraphics[width=10cm]{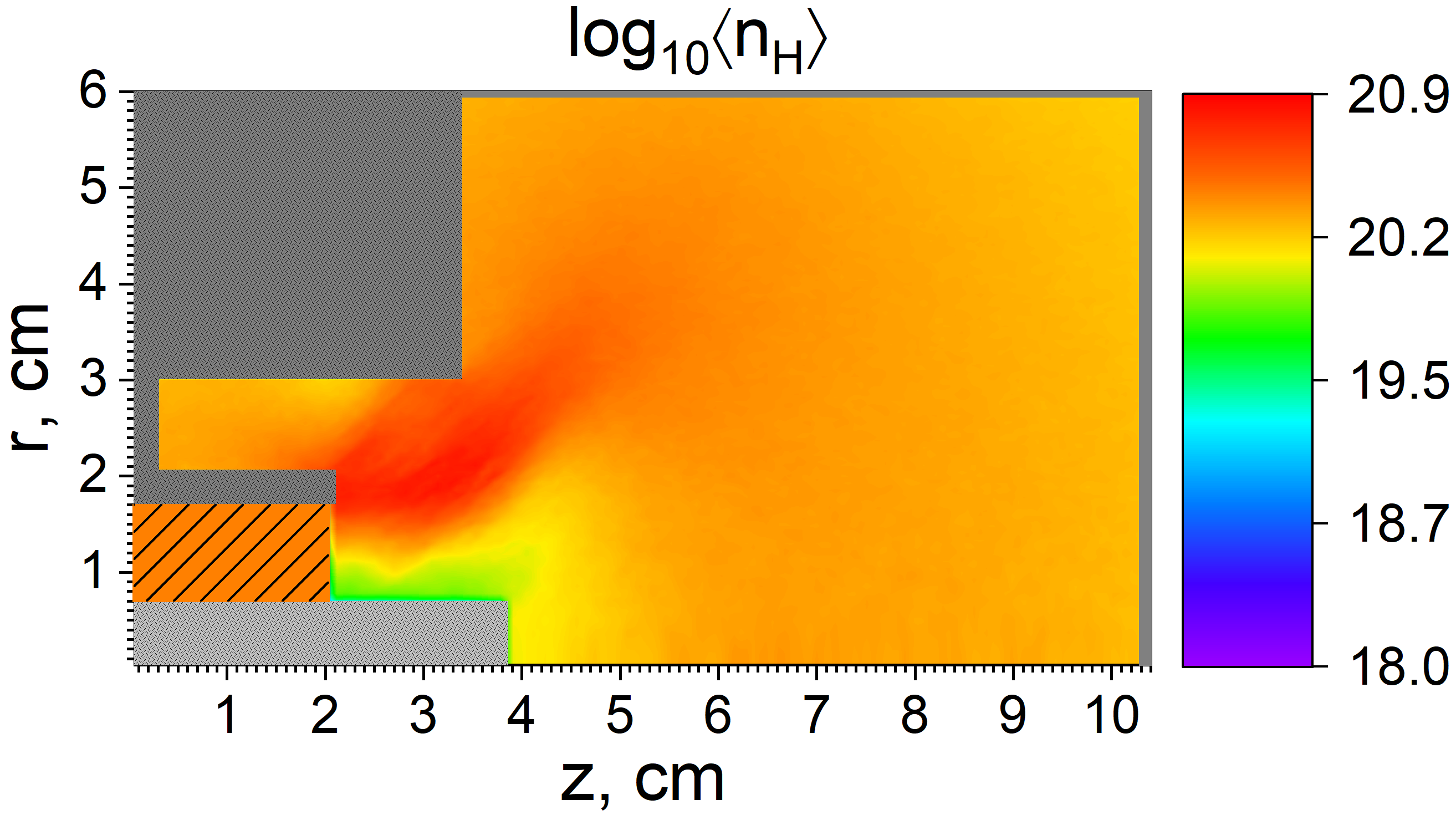}
\caption{Logarithm of the atomic hydrogen density (measured in m$^{-3}$).}
\label{Fig1H}
\end{figure}

Finally, we address the atomic hydrogen species. The corresponding profile
is a result of the production dominated by the molecular hydrogen dissociation reaction R17 and the transport processes. 
The period-averaged dissociation rate is shown in Fig.\ref{Fig2H} and can be linked to the energetic electron populations shown in Fig.\ref{denEnEl}, left. Evidently,
the population close to the central rod is more intense. This can be explained if the electron heating processes taking place there (see the discussion above) generate
electrons with higher energies lying in the ``superhot'' part of the energetic tail. The dissociation is assumed
to produce an isotropic distribution of hydrogen atoms in the center-of-mass frame with $1.73$ eV of kinetic energy,  
yielding fast moving neutral particles. The latter reflect from the stainless steel surfaces with $90 \%$ and from the alumina surface with $99.9 \%$ probability
along with getting elastically scattered on hydrogen molecules, which generates a non-trivial distribution of the hydrogen radical density, see Fig.\ref{Fig1H}.

\begin{figure}[h]
\centering
\includegraphics[width=7cm]{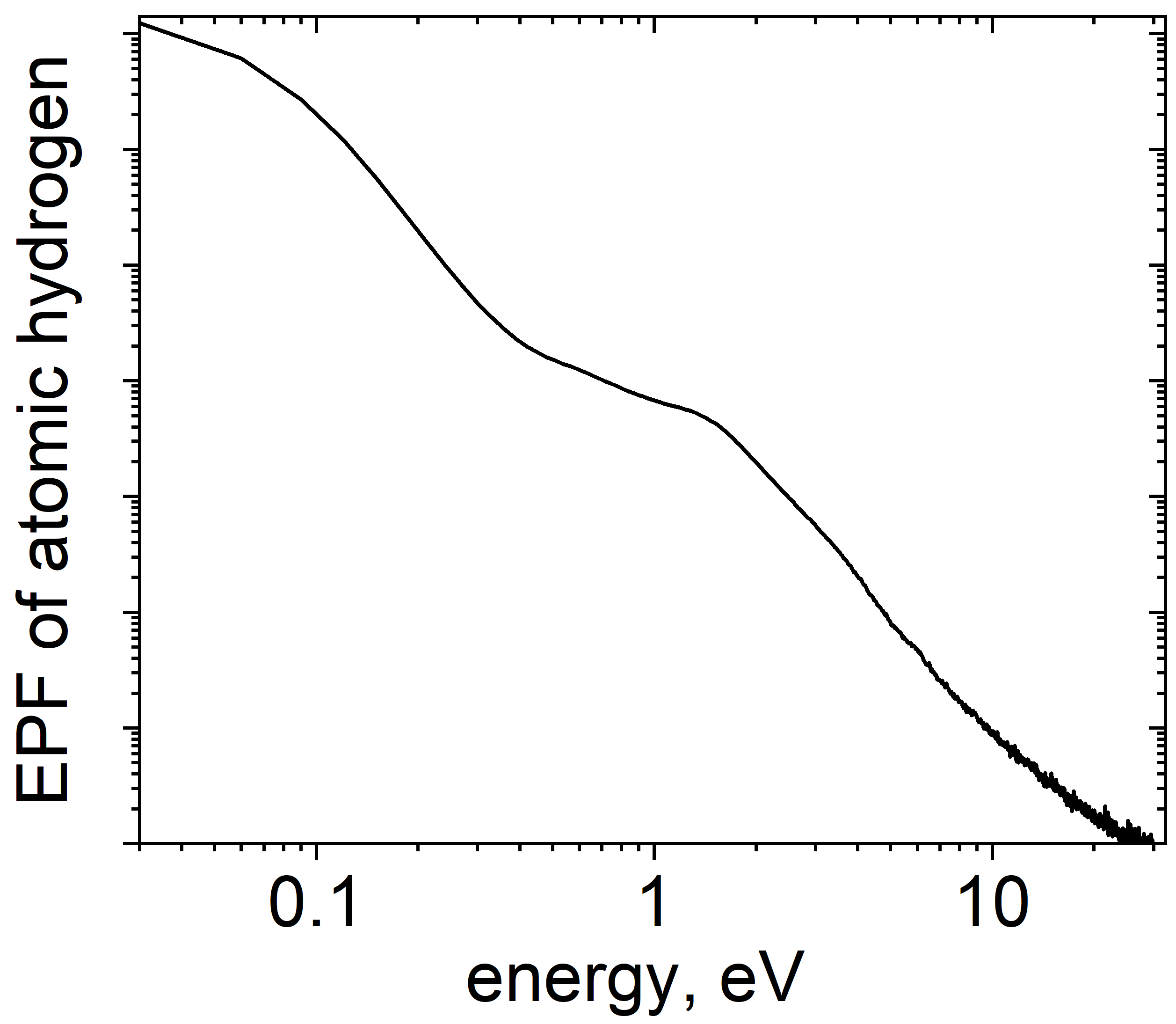}
\caption{Energy probability function of the atomic hydrogen plotted on the logarithmic scale.}
\label{NEPF_125W}
\end{figure}

To assess the significance of potential interaction between hydrogen radicals and reactor environment it is also
necessary to know energy distribution of the former. Fig.\ref{Fig2H} demonstrates that the corresponding energy
probability function features three groups of particles. The mid-energy group forms a quasi-plateau resulting from 
the hydrogen atoms produced with the energy $1.73$ eV in the dissociation reaction, whereupon they gradually loose
energy due to the elastic collisions. Due to the energy dependence of the elastic scattering cross-section 
the low-energy radicals have a higher slope compared to the mid-energy group. The high-energy group can be
explained by hydrogen atoms produced from ion species via various charge-transfer-like reactions (reactions R24, R25, R28, R30, R31, and R33).
Note that even in the low-energy group there is a large amount of relatively hot hydrogen radicals compared to the temperature
of the background molecular hydrogen (0.3 eV).

%%%%%%%%%%%%%%%%%%%%%%
%\clearpage
%%%%%%%%%%%%%%%%%%%%%%
\section{Conclusions} \label{sec4}

The present work describes a numerical model based on the fully electromagnetic charge- and energy-conserving implicit particle-in-cell/Monte Carlo approach
implemented in the code ECCOPIC2M-M developed at the Ruhr University Bochum. The model aims at describing an electron cyclotron resonance discharge in molecular hydrogen operated at a low pressure. 
Compared to a previous work using the electromagnetic charge- and energy-conserving implicit PIC \cite{Eremin2023}, the present model features an external magnetic field, a relatively complex chemistry and adaptive particle management. The code output is validated with experimental data for
electron density, temperature, and molecular hydrogen dissociation fraction previously obtained in \cite{Bogaers2021}. In general, the ECCOPIC2M-M results
demonstrate better agreement with the experimental data compared to a fluid-based code COMSOL, especially when kinetic plasma characteristics such as the electron
energy distribution are considered. This is related to the fact that the latter is strongly non-Maxwellian, exhibiting up to three different electron groups in case
of the PIC simulations. The PIC simulations show that the generation of various groups of energetic electrons can be explained by a number of additional electron heating mechanisms in addition to the ECR heating.
Whereas electrons are heated at the fundamental ECR harmonic in the discharge bulk plasma, other electron heating mechanisms take place close to the conducting surfaces. ECR-heated energetic electrons
with large pitch angles are efficiently confined by the background magnetic field created by permanent magnets. The atomic hydrogen fraction demonstrates a nonuniform density profile as a result of a combination of the nonuniform production via the dissociation of molecular hydrogen by electron impact and complex kinetic transport processes at a relatively large Knudsen number, involving reflections from geometrical features of the reactor. Due to the pronouncedly non-Maxwellian electron energy distribution, the ECCOPIC2M-M predicts a much higher dissociation of molecular hydrogen in comparison to COMSOL, in good agreement with experimental observations. Finally, the energy distribution of hydrogen atoms exhibits the presence of a relatively large amount of energetic particles, which can mainly be attributed to production of 1.73 eV hydrogen atoms resulting from the dissociation reaction, whereupon they become decelerated in elastic scattering events. The characteristics of such energetic hydrogen radicals are important in the context of their potential interaction with the reactor environment.

\section*{Acknowledgments}

This work was conducted in the framework of the RUB-ASML project ``ECR PIC Computational Plasma Investigation''.

\begin{appendix}

\section{calculation of the external magnetic field generated by cylindrical permanent magnets}

According to \cite{Latrasse2017}, the magnetic field in Aura-Wave source is created by two 
axially magnetized permanent magnets embedded in the central metal rod. The magnets are
mounted in opposition (with the magnet at the tip of the rod looking
into the chamber magnetized in the negative axial direction and the other magnet magnetized
in the positive direction). The corresponding magnetic field can be calculated using
an approach proposed in \cite{Reich2016}, which provides

\begin{equation}
\begin{array}{lll}
B_r(r,z) = B_{r0} \sum\limits^2_{i=1}\sum\limits^2_{n=1} (-1)^{i+n} R_n \int\limits^{2\pi}_0 d\phi^\prime \int\limits^{z_{i,max}}_{z_{i,min}} dz^\prime
\frac{(z-z^\prime)\cos(\phi^\prime)}{(r^2+R_n^2+(z-z^\prime)^2-2rR_n\cos(\phi^\prime))^{3/2}} && \\
B_z(r,z) = B_{r0} \sum\limits^2_{i=1}\sum\limits^2_{n=1} (-1)^{i+n} R_n \int\limits^{2\pi}_0 d\phi^\prime \int\limits^{z_{i,max}}_{z_{i,min}} dz^\prime
\frac{R_n-r\cos(\phi^\prime)}{(r^2+R_n^2+(z-z^\prime)^2-2rR_n\cos(\phi^\prime))^{3/2}}, &&
\end{array}
\end{equation}
where $B_{r0}$ is the remnance (taken to be equal to $1.02$ T), $R_n=\{6,1.5\}$ mm for $n=\{1,2\}$ along with $z_{i,min}=\{5,20\}$ mm and $z_{i,max}=\{20,35\}$ mm
for $i=\{1,2\}$, respectively.

\end{appendix}

%\appendix
%\section{Magnetic field approximation} \label{Appendix_a}

\section*{References}
\bibliographystyle{ieeetr}
%\bibliographystyle{unsrt}
%\begin{thebibliography}{99}

\bibliography{references}
%\end{thebibliography}

\end{document}